\documentclass{JAC2003}

\usepackage{graphicx}

\begin{document}
\title{KLOE Results on Rare $K^0$ Decays}

\author{KLOE Collaboration\thanks{\small{A.~Aloisio, F.~Ambrosino,
A.~Antonelli, M.~Antonelli, C.~Bacci, G.~Bencivenni, S.~Bertolucci,
C.~Bini, C.~Bloise,
V.Bocci, F.~Bossi, P.~Branchini, S.~A.~Bulychjov,
R.~Caloi, P.~Campana, G.~Capon, T.~Capussela, G.~Carboni, F.~Ceradini,
F.Cervelli, F.~Cevenini, G.~Chiefari, P.~Ciambrone, S.~Conetti,
E.De~Lucia, P.~De~Simone, G.~De~Zorzi, S.~Dell'Agnello,
A.Denig, A.~Di~Domenico, C.~Di~Donato, S.~Di~Falco, B.~Di~Micco,
A.~Doria,M.~Dreucci, O.Erriquez, A.Farilla, G.~Felici, A.~Ferrari,
M.~L.~Ferrer, G.~Finocchiaro, C.~Forti, P.~Franzini, C.~Gatti,
P.Gauzzi, S.~Giovannella, E.~Gorini, E.Graziani, M.~Incagli, W.~Kluge,
V.~Kulikov, F.~Lacava, G.~Lanfranchi, J.~Lee-Franzini, D.~Leone,
F.Lu, M.~Martemianov, M.~Matsyuk, W.Mei, L.~Merola, R.~Messi,
S.~Miscetti, M.~Moulson, S.~M\"uller, F.Murtas, M.~Napolitano,
F.~Nguyen, M.~Palutan, E.Pasqualucci, L.~Passalacqua, A.~Passeri,
V.~Patera, F.~Perfetto, E.~Petrolo, L.~Pontecorvo, M.~Primavera,
P.~Santangelo, E.~Santovetti, G.~Saracino, R.~D.~Schamberger,
B.~Sciascia, A.~Sciubba, F.~Scuri, I.~Sfiligoi, A.~Sibidanov,
T.~Spadaro, E.~Spiriti, M.~Tabidze, M.~Testa, L.~Tortora, P.~Valente,
B.~Valeriani, G.~Venanzoni, S.~Veneziano, A.~Ventura, S.~Ventura,
R.~Versaci, I.~Villella, G.~Xu}}
\\
Presented by S. Dell'Agnello\thanks{E-mail:
simone.dellagnello@lnf.infn.it}, INFN-LNF, Frascati (RM), 00044, ITALY}

\maketitle

\begin{abstract}

This paper describes the neutral Kaon dataset and physics
measurements of KLOE at DA$\Phi$NE. A brief discussion on
the prospects for detector upgrades and analysis
of other $K_S$ rare decays is also included.

\end{abstract}

\section{$K^0$ DATASET}

The KLOE detector is described in detail elsewhere \cite{NIMs}.
The experiment has integrated a luminosity of 25/170/280
pb$^{-1}$ in the years 2000/1/2, corresponding to 27/180/296
$\cdot$10$^6$ produced $K_L K_S$ events. The highest instantaneous
luminosity has been 8 $\cdot$ 10$^{31}$ cm$^{-2}$ sec${-1}$, while
the highest daily luminosity has been 4.5 pb$^{-1}$.

\section{BENCHMARK $K^0$ MEASUREMENTS}

KLOE has started its $K^0$ physics program with benchmark
measurements which have firmly established its capability in
this field. Most analysis share these features:
\begin{itemize}
\item Request a $K_L$ or $K_S$ tag. The tagging efficiency is
measured from data control samples (an ultimate 0.1\% accuracy
is expected for the full 0.5 fb$^{-1}$ luminosity).
\item Trigger efficiency is measured mainly from the data.
\item $\sqrt{s}$, and $e^+ e^-$ interaction point (IP) are measured online
run-by-run with Bhabha scattering events. Typical accuracies for 0.1
pb$^{-1}$ runs are: $\sigma(\sqrt{s}) \sim$ 40 keV, $\sigma(P_{\Phi})
\sim$ 30 keV, $\sigma(X_{IP}) \sim \sigma(Y_{IP})
\sim$ 30 $\mu$m.
\item $\phi$ production time, $T_{\phi}$, are reconstructed offline with
accuracy $\sigma(T_{\phi}) \sim$ 50 psec.
\item Monte Carlo simulation (MC) are used mainly for acceptance and
geometry corrections. The MC photon and track reconstruction
efficiencies are properly scaled to data control samples.
\end{itemize}

An important measurement performed by KLOE is the ratio
of the branching ratios (BR) of $K_S \to \pi \pi$ decays.

\subsection{$K_S \to \pi \pi$ Decays}

\begin{figure}[hb]
\centering
\includegraphics*[width=71mm]{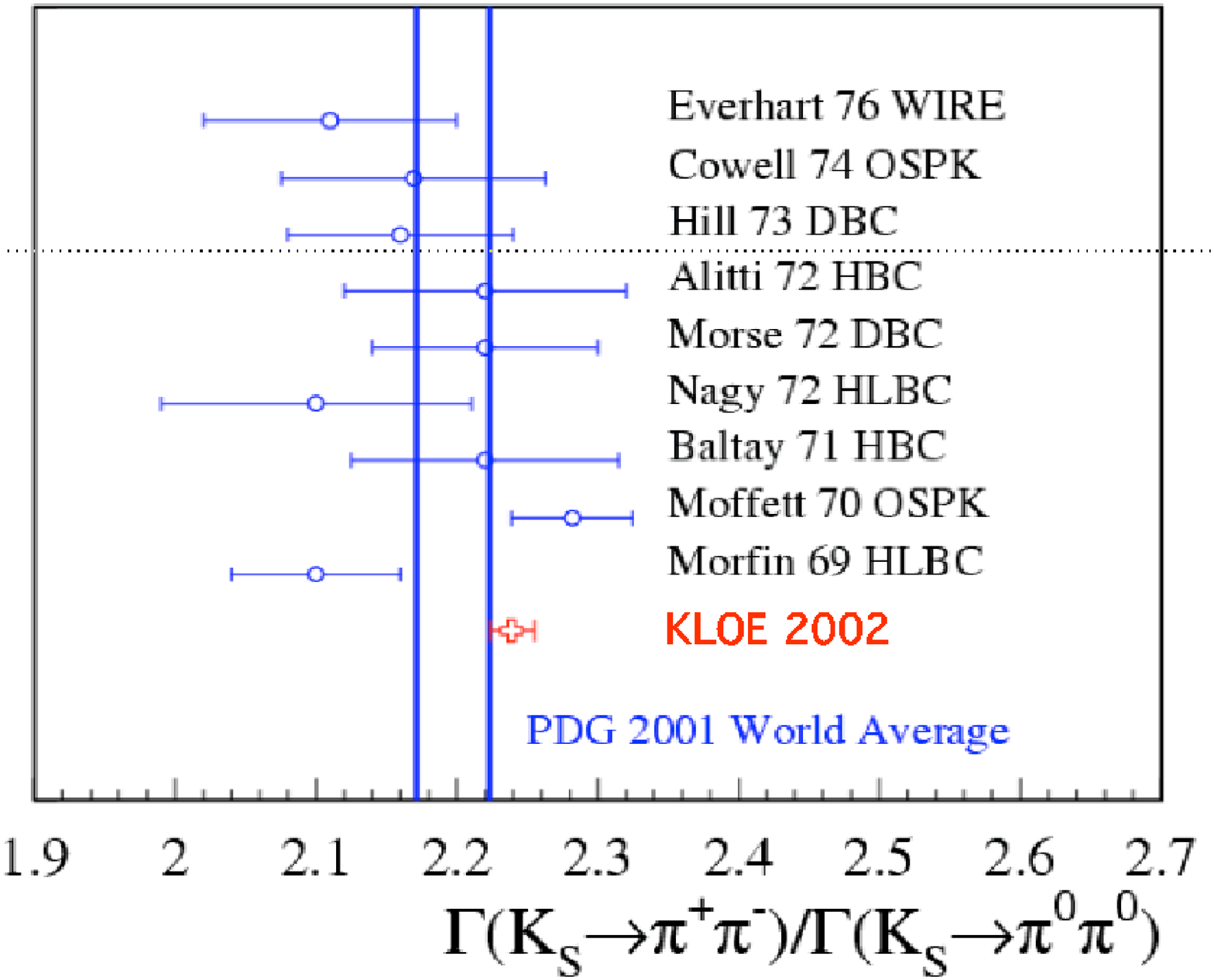}
\caption{KLOE measurement performed with 17 pb$^{-1}$
(1.1$\cdot$10$^6$/0.8$\cdot$10$^6$ tagged
$K_S\to \pi^+ \pi^-$/$K_S\to \pi^0 \pi^0$)
compared to previous results and to the PDG average.}
\label{Ks2pifig}
\end{figure}

$K_S$ decays are tagged via the reconstruction of $K_L$ interactions
in the EM calorimeter. The measurement of
$\Gamma(K_S\to\pi^+\pi^-(\gamma))$/$\Gamma(K_S\to\pi^0\pi^0)$
is a benchmark for the measurement of $\epsilon$'/
$\epsilon$ via the double ratio method and it can shed light on
the $\Delta I=1/2$ suppression rule. It also gives information on
the values of the strong phase shifts ($\delta_0 - \delta_2$) and
the electromagnetic isospin breaking ($\gamma_0 - \gamma_2$). The KLOE
result is fully inclusive of $\pi^+\pi^-\gamma$ final states, it has an
unprecedented statistical accuracy of $\sim$0.1\% and is limited
at present by
the systematic uncertainty which is estimated primarily from the data.
$\Gamma(K_S \to \pi^+\pi^-(\gamma))$/$\Gamma(K_S \to \pi^0\pi^0)$ =
2.236 $\pm$ 0.003(stat) $\pm$ 0.015(syst) \cite{Ks2pi} (see Figure
\ref{Ks2pifig}). The present 0.7\% systematic error is expected to scale
down to below 0.2\% for 0.5 fb$^{-1}$. From this measurement we extract
$\chi_0 - \chi_2$ = (48 $\pm$ 3)$^{\circ}$ ($\chi_I =
\delta_I + \gamma_I$), which is to be compared to the
estimate from the PDG widths of
$\chi_0 - \chi_2$ = (56 $\pm$ 8)$^{\circ}$ \cite{cirigliano}.
Other estimates are:
$\delta_0 - \delta_2$ = (45 $\pm$ 6)$^{\circ}$ \cite{chipt}
from chiral perturbation theory ($\chi$PT) and
$\delta_0 - \delta_2$ = (47.7 $\pm$ 1.5)$^{\circ}$ \cite{pipiscatt}
from $\pi\pi$ scattering.

\subsection{Invariant Mass and Lifetimes}

\begin{figure}[ht]
\centering
\includegraphics*[width=55mm]{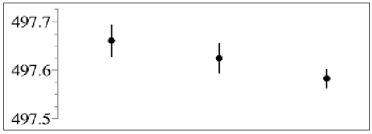}
\caption{The $K_S$ mass as measured by CMD-2 (left data point),
NA48 (center) and KLOE (right, see \cite{ksmass}).}
\label{ksmassfig}
\end{figure}

\begin{figure}[ht]
\centering
\includegraphics*[width=65mm]{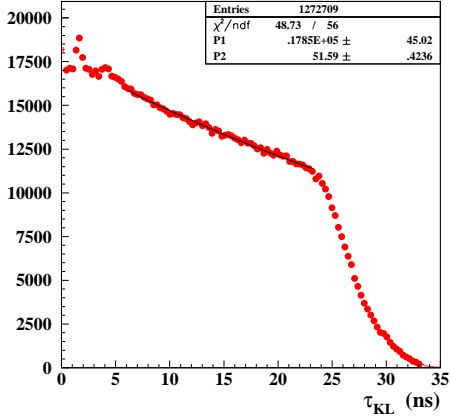}
\caption{The KLOE $K_L \to 3 \pi^0$ lifetime (statistical error only):
(51.6 $\pm$ 0.4) nsec. PDG: (51.7 $\pm$ 0.4) nsec \cite{tklfig}.}
\label{tklfig}
\end{figure}

KLOE has also performed preliminary measurements of the $K_S$ invariant
mass ($K_S \to \pi^+ \pi^-$ channel) and of the $K_L$ lifetime,
reaching accuracies which are better or comparable to previous measurements
(see figures \ref{ksmassfig} and \ref{tklfig}).

\begin{figure}[ht]
\centering
\includegraphics*[width=62mm]{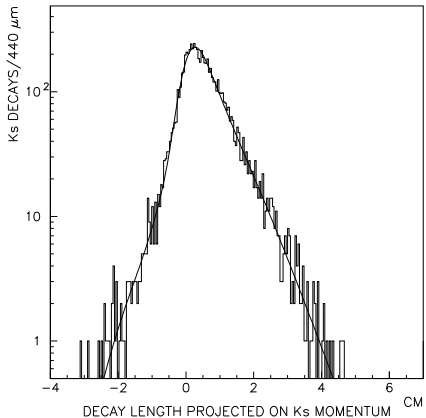}
\caption{The $K_S \to \pi^+ \pi^-$ decay length distribution \cite{tksfig}.}
\label{tksfig}
\end{figure}

The $K_S \to \pi^+ \pi^-$ lifetime has also been studied and found
to be consistent with the PDG average (see fig. \ref{tksfig}). The
vertex resolution is $\sim$1/3 of the decay length ($\sim$6 mm).

\section{BR($K_L \to \gamma \gamma$)}

\begin{figure}[ht]
\centering
\includegraphics*[width=75mm]{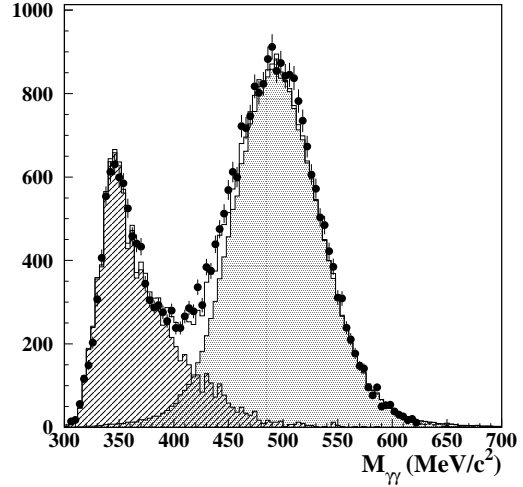}
\caption{$\gamma \gamma$ invariant mass spectra: data (points)
and Monte Carlo (histogram) used to measure BR($K_L \to \gamma \gamma$);
signal is at the right peak.}
\label{klggfig}
\end{figure}

This BR has a large long-distance contribution via the pseudoscalar
mesons ($\pi^0$, $\eta$, $\eta'$). It can be calculated in $\chi$PT
and is sensitive to the pseudoscalar meson mixing angle, $\theta_P$.
Its value enters and dominates the long-distance contribution in
$K_L \to \mu^+ \mu^-$. The measurement reported here is based on 362
pb$^{-1}$ and $1.6\times10^8$ $K_L$ decays tagged by observing
$K_S \to \pi^+ \pi^-$ \cite{klgg}. The large $K_L \to 3\pi^0$ background
is suppressed by exploiting the 2-body kinematics, yielding 27375
estimated signal events (see figure \ref{klggfig}). We measure the ratio
$\Gamma(K_L \to \gamma \gamma)/\Gamma(K_L \to 3\pi^0)$ = 
(2.79 $\pm$ 0.02 $\pm$ 0.02)$\times$10$^{-3}$. The NA48 measurement
is (2.81 $\pm$ 0.01 $\pm$ 0.02)$\times$10$^{-3}$
\cite{klggNA48}. Using the PDG value of BR($K_L \to 3\pi^0$), KLOE
gets BR($K_L \to\gamma \gamma$) = (5.89 $\pm$ 0.07$_{stat \oplus syst}$
$\pm$ 0.08$_{BR(K_L \to 3\pi^0)}$)$\times$10$^{-4}$. This value of
the BR is in agreement with $\chi$PT, if $\theta_P$ is close to the KLOE
measurement $\theta_P$ = (-12.9$^{+1.9}_{-1.6}$)$^{\circ}$ \cite{tetap}.

\section{BR($K_S \to \pi \lowercase{e} \nu$)}

\begin{figure}[htb]
\centering
\includegraphics*[width=75mm]{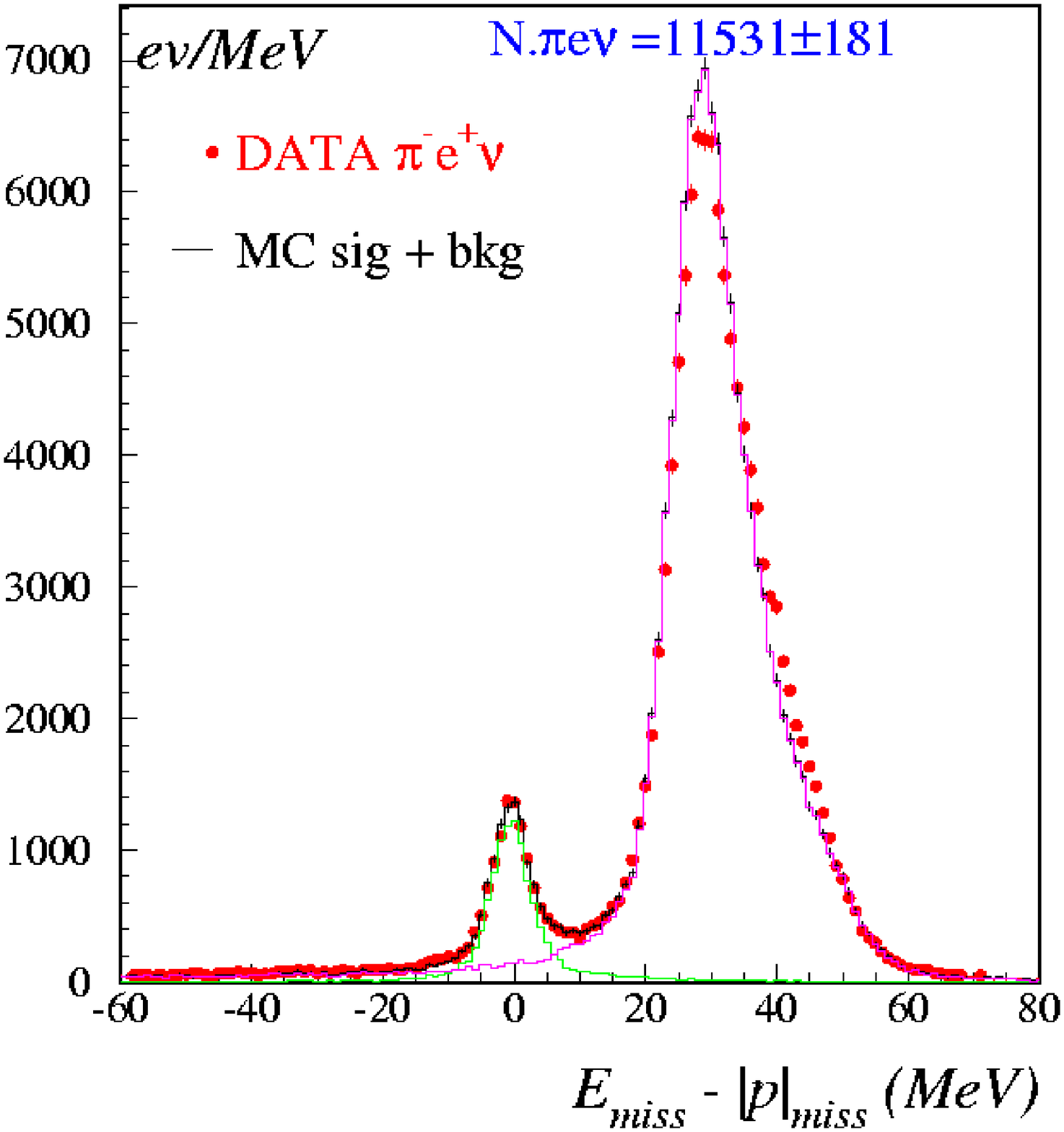}
\caption{$K_S \to \pi^+ e^- \bar{\nu}$ signal (peak at 0) and background
(mostly $\pi$ decays in flight) in $\sim$ 0.5 fb$^{-1}$.}
\label{ks_eplusfig}
\end{figure}

\begin{figure}[ht]
\centering
\includegraphics*[width=75mm]{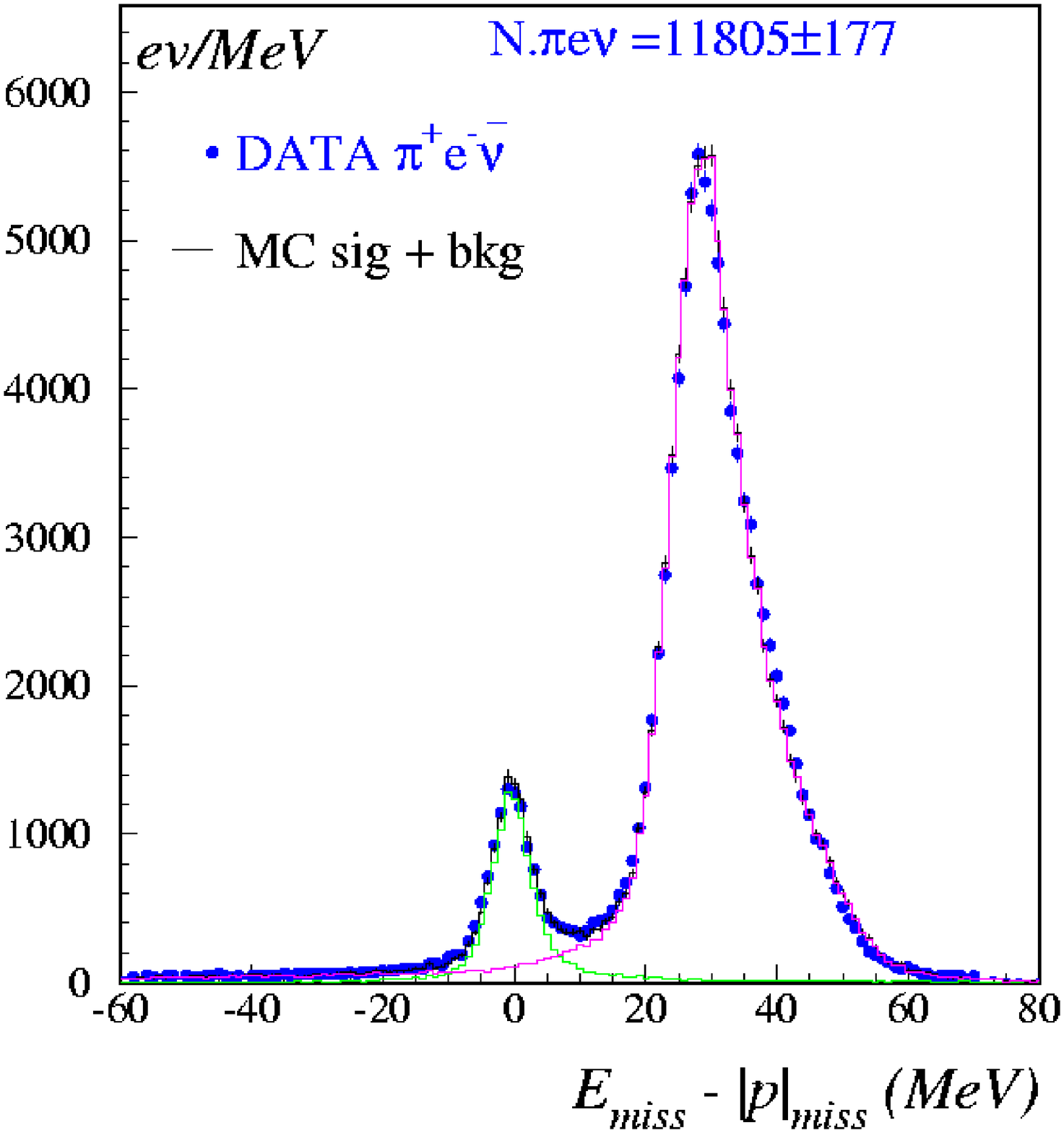}
\caption{$K_S \to \pi^- e^+ \nu$ signal (peak at 0) and background
(mostly $\pi$ decays in flight) in $\sim$ 0.5 fb$^{-1}$.}
\label{ks_eminusfig}
\end{figure}

$K_S$ decays are tagged via reconstruction of $K_L$ interactions
in the EM calorimeter. The $K_S \to \pi^+ \pi^-$ background, which
is 10$^3$ times larger, is suppressed by exploiting the 3-body
kinematics and the $e$/$\pi$ discrimination based on time-of-flight
measurement with the calorimeter. The remaining background ($\sim$5\%)
is given by $K_S \to \pi^+ \pi^-$ with $\pi \to \mu \nu$
inside the drift chamber. The data is fit to the sum of signal and
background MC shapes and normalized to the KLOE-measured
BR($K_S \to \pi^+ \pi^-$) to get the signal branching ratio.
Using 170 pb$^{-1}$ of 2001 data we report the preliminary
measurement (which supersede our first published result obtained
with 17 pb$^{-1}$ \cite{kssemil17pb}):
\begin{itemize}
\item BR($K_S \to \pi^{\mp} e^{\pm} \nu (\bar{\nu})$) = (6.81 $\pm$ 0.12 $\pm$ 0.10)
      $\cdot$ 10$^{-4}$
\item BR($K_S \to \pi^- e^+ \nu$) = (3.46 $\pm$ 0.09 $\pm$ 0.06)
      $\cdot$ 10$^{-4}$
\item BR($K_S \to \pi^+ e^- \bar{\nu}$) = (3.33 $\pm$ 0.08 $\pm$ 0.05)
      $\cdot$ 10$^{-4}$.
\end{itemize}
These results agree with the Standard Model expectations and improves
significantly the previous CMD-2 measurement: (7.2 $\pm$ 1.4) $\cdot$ 10$^{-4}$
\cite{CMD-2}.
Work is in progress to take into account the contribution of radiative
$K_S$ semileptonic decays, to employ an improved version of the MC.
Figures \ref{ks_eplusfig} and \ref{ks_eminusfig} show preliminary
estimates of the number of signal events in 0.5 fb$^{-1}$.

\subsection{$K_S$ Charge Asymmetry and Test of CPT}

The charge asymmetry of the $K_S$ semileptonic decay, $A_S$, has been
measured for the first time by KLOE. The preliminary result for
170 pb$^{-1}$ is:
\begin{equation}
A_S = (19 \pm 17 \pm 6) \cdot 10^{-3}.
\end{equation}
A non-zero difference between $A_S$ and the $K_L$ semileptonic asymmetry,
$A_L$, would imply the violation of the CPT conservation law:
\begin{equation}
A_S - A_L = 4 \; Re(\delta_K - D), 
\end{equation}
where: $\delta_K$ is the CPT violation in the $K^0$ mixing and
$D$ is a term containing the $K^0$ semileptonic decay amplitudes
of the weak Hamiltonian, which violates both CPT and the
$\Delta S$ = $\Delta Q$ rule.
The world-average value of $A_L$ is (3.322 $\pm$ 0.055)$\cdot$10$^{-3}$.
For comparison: the CPLEAR measurement of CPT violation is \cite{cplear}
$Re(\delta_K)$ = (2.9 $\pm$ 2.7)$\cdot$10$^{-4}$
if $\Delta S$ = $\Delta Q$ is assumed and 
$Re(\delta_K)$ = (3.0 $\pm$ 3.4)$\cdot$10$^{-4}$
if $\Delta S$ = $\Delta Q$ is not assumed.

\subsection{Test of $\Delta S = \Delta Q$ Rule}

In the Standard Model the are no $\Delta S \neq \Delta Q$ transitions
at the lowest order. Such transitions are described by the phenomenological
parameters $x$ ($\Delta S \neq \Delta Q$ in the $\bar{K}^0$ decay to
$e^+$) and $\bar{x}$ ($\Delta S \neq \Delta Q$ in the $K^0$ decay to
$e^-$), which contain the $K^0$ semileptonic amplitudes. The combination
$x_+$ = ($x$ + $\bar{x}$)/2 describes the amount of $\Delta S \neq \Delta Q$
when CPT is conserved. $x_+$ can be expressed as:
\begin{equation}
{BR(K_S \to \pi e \nu)/\tau_S \over BR(K_L \to \pi e \nu)/\tau_L}
        = {\Gamma_S^{semil} \over \Gamma_L^{semil}} = 1 + 4 \; Re(x_+).
\end{equation}
The value measured by KLOE with 170 pb$^{-1}$ of 2001 data is:
$Re(x_+)$ = $(3.3 \pm 5.2 \pm 3.5)\cdot 10^{-3}$,
which is to becompared with
$Re(x_+)$ = $(-1.8 \pm 4.1 \pm 4.5)\cdot 10^{-3}$ (and
$Im(x_+)$ = $(1.2 \pm 1.9 \pm 0.9)\cdot 10^{-3}$) by CPLEAR \cite{cplear}.
An update of this meaasurement with 0.5 fb$^{-1}$ will follow soon.

\section{MEASUREMENT OF $V_{US}$}

KLOE is giving a significant  contribution to the measurement
of the $V_{us}$ CKM matrix element, which constrains
the internal consistency of the Standard Model. The experimental
inputs to extract $V_{us}$ from the Kaon semileptonic
decays are BRs and lifetimes. The theoretical inputs are the
linear slopes of the $K_{\ell 3}$ form factors, $\lambda_{\pm}$,
and the values of these form factors at zero 4-momentum transfer
to the leptons, $f_{\pm}(0)$. The uncertainty on $V_{us}$
for $K_{e3}$ (for which $f_{-}$ is negligible) depends on these inputs
as:
\begin{equation}
{\delta V_{us} \over V_{us}} = {1 \over 2} {\delta BR \over BR}
\oplus {1 \over 2} {\delta \tau \over \tau}
\oplus {1 \over 20} {\delta \lambda_+ \over \lambda_+}
\oplus {\delta f_+(0) \over f_+(0)}.
\end{equation}

\begin{figure}[ht]
\centering
\includegraphics*[width=80mm]{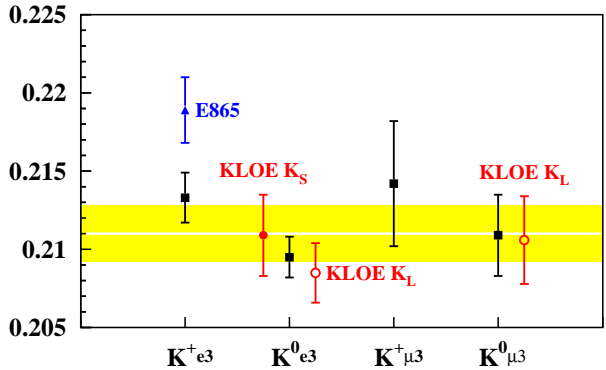}
\caption{$V_{us} \cdot f_+(0)$ vs semileptonic Kaon decay
mode from old (black points) and the recent E865 measurement.
The KLOE preliminary results are also indicated.}
\label{vusfig}
\end{figure}

In addition to the above mentioned $K_S$
semileptonic BR, KLOE has also preliminary measurements of
the $K_L$ BRs in both the electron and muon channel, based
on a statistics of 78 pb$^{-1}$. Our measurements of
$V_{us} \cdot f_+(0)$ are consistent with old $K^0_{e3}$ and
$K^0_{\mu3}$ measurements and are lower than the recent E865
result \cite{E865}, as shown in figure \ref{vusfig}.

KLOE will have BR($K^+_{e3}$) soon and plans on measuring BRs
with $\ll$1\% accuracy. KLOE can also improve the measurements
of $\lambda_+$, $\lambda_0$, $\tau_L$ and $\tau_S$.

\section{Conclusions and Prospects}

Current $K^0$ physics program is driven significantly by
statistics (0.5 fb$^{-1}$) and, to a lesser extent, by the
DA$\Phi$NE background level (which, however, has been
constantly decreasing through the years). The MC has been
vastly improved: (i) background events from data are
injected into simulated events on a run-by-run basis;
(ii) the detector/trigger response, materials and
geometry have been refined, (iii) generators for radiative $K^0$
decays have been added. The new MC allows us to tackle the
next $K_S$ rare decays. For example, with 0.5 fb$^{-1}$ we
set the preliminary limit BR($K_S \to 3\pi^0)< 2.2 \cdot 10^{-7}$,
90\% CL (5 events on a background of 3.1 $\pm$ 1.9),
the best limit to date.

The results on BR($K_S \to \pi e \nu$), BR($K_L \to \gamma \gamma$)
and BR($K_S \to 3\pi^0$) show that KLOE can measure
rare decays. A factor 10 luminosity increase is
needed to measure the next more rare $K_S$ decays and
$\epsilon'/\epsilon$.

In the prospect of a long term running of KLOE (and more than
5 fb$^{-1}$ statistics) some detector upgrades are strongly
advisable and beficial to the analysis.
\begin{itemize}
\item New and smaller KLOE-DA$\Phi$NE interaction region (IR).
As a consequence, new and smaller EM calorimeters around the
IR quadrupoles (QCAL) are necessary in order not to introduce
large dead regions into the detector volume).
\item With the new IR+QCAL there will be room for
a new compact inner vertex detector inside the drift chamber
(10 $<$ radius $<$ 25 cm), capable of measuring $also$ the 
longitudinal coordinate. This detectod will have the following
features:
\begin{itemize} 
\item[-] help the track pattern recognition at small
radii where beam background is more severe;
\item[-] improve vertexing at IP and interferometry for all-charged
events, like $K_S K_L \to \pi^+ \pi^- \pi^+ \pi^-$,
$K_S K_L \to \pi^+ \pi^- \pi e \nu$ and $K_S K_L \to \pi e \nu \pi e \nu$;
\item[-] help the identification of Kaon interactions in the
drift chamber inner wall (esp. Q-exchange);
\item[-] the beam pipe at IP will be made of pure Beryllium,
to ease complexity in offline event reconstruction; the current
spherical geometry can be abandoned in favor of an easier
machinable shape.
\end{itemize}
\item The readout granularity of existing calorimeter will
be increased by removing light guides and phototubes, to be
replaced by smaller light-collections elements. This would improve
the energy clustering algorithm and enhance particle ID (PID)
\end{itemize}

The specific choices for the new inner vertex detector and
calorimeter readout devices are to be studied. It should
be noted that in 2002 KLOE has completed its first
successful detector upgrade: the drift chamber has been
instrumented with ADCs to help PID by means of dE/dx, which
is effective especially for charged Kaon physics.

KLOE expects to collect $\sim$ 2 fb$^{-1}$ in the physics run
starting in spring 2004, which should benefit from the upgrades
performed by DA$\Phi$NE in 2003, which included a new IR with
simplified optics.

\end{document}